\documentclass[aps,prl,twocolumn,amsmath,amssymb]{revtex4}

\usepackage{graphicx}
\usepackage{dcolumn}
\usepackage{bm}
\usepackage{epsfig}
\usepackage{subfig}
\usepackage{amsmath}
\usepackage{amsfonts}

\newcommand{\ba}{\begin{eqnarray}}
\newcommand{\ea}{\end{eqnarray}}
\newcommand{\be}{\begin{equation}}
\newcommand{\ee}{\end{equation}}
\newcommand{\bea}{\begin{eqnarray}}
\newcommand{\eea}{\end{eqnarray}}

\usepackage{theorem}

\theorembodyfont{\rmfamily}

\theoremstyle{break}

\def\QED{~\rule[-1pt]{5pt}{5pt}\par\medskip}

\begin{document}


\title{Sequential feedback scheme outperforms the parallel scheme for Hamiltonian parameter estimation}
\author{Haidong Yuan}
\email{hdyuan@mae.cuhk.edu.hk}
\affiliation{Department of Mechanical and Automation Engineering, The Chinese University of Hong Kong, Shatin, Hong Kong}


\date{\today}

\begin{abstract}
Measurement and estimation of parameters are essential for science and engineering, where the main quest is to find out the highest achievable precision with given resources and design schemes to attain it. Two schemes, the sequential feedback scheme and the parallel scheme, are usually studied in quantum parameter estimation. While the sequential feedback scheme represents the most general scheme, it remains unknown whether it can outperform the parallel scheme for any quantum estimation tasks. In this Letter we show that the sequential feedback scheme has a 3-fold improvement over the parallel scheme for Hamiltonian parameter estimations on 2-dimensional systems, and an order of $O(d+1)$ improvement for Hamiltonian parameter estimation on $d-$dimensional systems. We also show that, contrary to the conventional belief, it is possible to simultaneously achieve the highest precision for estimating all three components of a magnetic field, which sets a benchmark on the local precision limit for the estimation of a magnetic field.
\end{abstract}
\maketitle
A pivotal task in science and technology is to find out the highest achievable precision in measuring and estimating parameters of interest with given resources and design schemes to reach that precision\cite{Giovannetti2011,GIOV04,GIOV06,Fujiwara2008,Escher2011,Tsang2013,Rafal2012,Knysh2014,Jan2013,Rafal2014,Alipour2014,Chin2012,Tsang2011,Berry2013,Berry2015,Berry2009,Berry2012,Hall2012,Tsang2012,HallPRX}. Typically to estimate some parameters $x=(x_1,x_2,\cdots,x_m)$ encoded in some dynamics $\phi_x$, a probe state $\rho_0$ is prepared which evolves under the dynamics $\rho_0\xrightarrow{\phi_x} \rho_x$. By performing Positive Operator Valued Measurements(POVM) $\{E_y\}$, on the output state $\rho_x$, one gets the measurement result $y$ with a probability $p(y|x)=Tr(E_y\rho_x)$. With a prior knowledge that the parameter is within some local interval, the variance of any unbiased estimator of $x$ is then bounded below by the Fisher information matrix $nCov(\hat{x})\geq I^{-1}(x)$\cite{HELS67,HOLE82,CRAM46,Rao}, where $n$ is the number of times that the procedure is repeated, $Cov(\hat{x})$ denotes the covariance matrix of the estimator, and $I(x)$ is the Fisher information matrix with the $ij$-th entry given by
$I_{ij}(x)=\int p(y|x)\frac{\partial lnp(y|x)}{\partial x_i}\frac{\partial lnp(y|x)}{\partial x_j} dy$\cite{Fisher}. The Fisher information matrix can be further bounded by the quantum Fisher information matrix(QFIM) $J(\rho_x)$, which gives the quantum Cram\'er-Rao bound\cite{HELS67, HOLE82,BRAU94,BRAU96}
$nCov(\hat{x})\geq I^{-1}(x)\geq J^{-1}(\rho_x).$

In multi-parameter estimation the quantum Cram\'er-Rao bound is usually not achievable even asymptotically\cite{HELS67, HOLE82,Genoni2013,vidrighin2014,crowley2014,humphreys2013,Yue2014,Zhang2014,Tsang2014,Kok2015,Spagnolo2012,Datta}. Two tradeoffs have to be considered in multi-parameter estimation: the first tradeoff is on the choice of measurements as the optimal measurements for different parameters are usually incompatible\cite{Gill2000}; the second tradeoff is on the choice of the probe states since the optimal probe states for different parameters are also usually different. These tradeoffs are usually dealt with by specifying a particular figure of merit taken as $Tr[Cov(\hat{x})G]$ with $G\geq 0$ then optimizing the measurements and probe states based on the figure of merit.

\begin{figure}
  \centering
  \includegraphics[width=\linewidth]{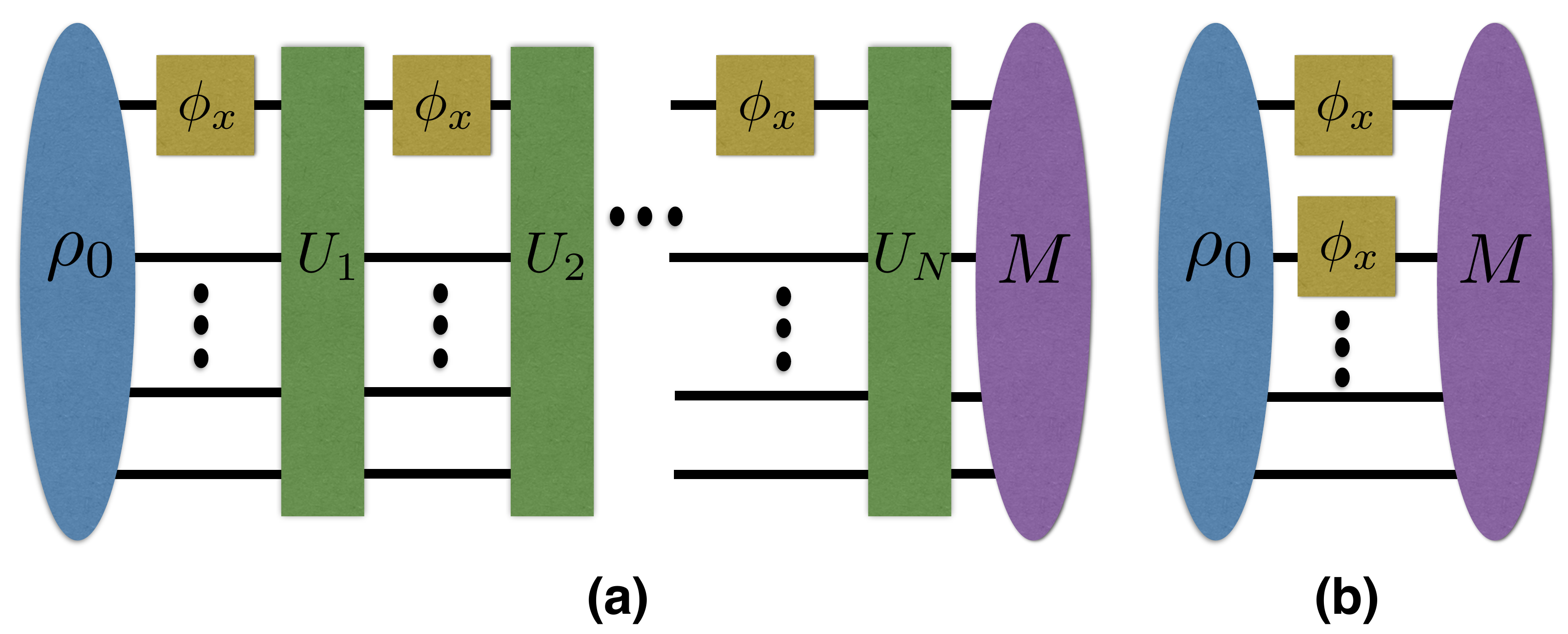}
  \caption{(a)Sequential feedback scheme.(b)Parallel scheme.}
  \label{fig:scheme}
\end{figure}

Besides the measurements and probe states, one also needs to optimize the schemes that arrange multiple uses of the dynamics to achieve the ultimate precision limit. Two schemes, the sequential feedback scheme and the parallel scheme, as shown in Fig.\ref{fig:scheme}, are usually studied. The sequential feedback scheme represents the most general scheme, which includes the parallel scheme as a special case when taking the controls as SWAP gates. Examples have been found in quantum channel discrimination that the sequential feedback scheme can outperform the parallel scheme for the discrimination of two quantum channels \cite{ChiribellaDP08,Harrow2010}.
In quantum parameter estimation it remains unknown whether the sequential feedback scheme can outperform the parallel scheme. Based on some upper bounds on the precision limit \cite{Escher2011,Rafal2014}, it has been shown that the sequential feedback scheme does not lead to higher precision in single-parameter quantum estimation under several dynamics, including the unitary\cite{GIOV06} and dephasing dynamics\cite{Rafal2012,Jan2013,Knysh2014, Rafal2014}. This has led to a conjecture that in the asymptotical limit the sequential feedback scheme provides no gains over the parallel scheme for quantum parameter estimation\cite{Rafal2014}.

In this Letter we show that the sequential feedback scheme outperforms the parallel scheme for Hamiltonian parameter estimation, here we focus on the estimations of small shifts of the parameters around some known values. We first study the estimation of the Hamiltonian for $SU(2)$ dynamics, which is a fundamental problem in quantum parameter estimation\cite{Peres2001,Acin01PRA,Chiribella2004,Bagan2004,Bagan20042,Chiribella2005,Fujiwara2001,Imai2007,Manuel2004,Manuel20042,Hayashi2006,Datta, pang2014,Liu2014} and closely related to the estimation of a magnetic field. It thus has many applications in quantum sensing, data storage, information processing and magnetic resonance, and has implications in quantum gyroscope, quantum reference frame alignments,etc\cite{Peres2001,Chiribella2004,Bagan2004,Bagan20042,Chiribella2005,Chiribella2014}. By considering the general sequential feedback scheme we obtain the ultimate local precision limit for the estimation of a magnetic field, and show that in this case the sequential feedback scheme outperforms the parallel scheme with a 3-fold improvement. We also show that, contrary to the conventional belief that some tradeoffs have to be made in the estimation of different parameters of a magnetic field, the optimal sequential feedback scheme achieves the highest precision for all three parameters of a magnetic field simultaneously, which sets a benchmark on the local precision of estimating multiple components of a magnetic field. We then show for the estimation of general Hamiltonians for $SU(d)$ dynamics the sequential feedback scheme outperforms the parallel scheme with an order of $O(d+1)$ improvement. This sheds light on the comparison between the two schemes. Extension to Hamiltonian parameter estimation with prior distributions are also discussed in the supplemental material. We also note that the sequential feedback scheme is more implementable for many current experimental settings since high-fidelity controls on small systems can now be routinely done while accurately preparing entangled states with many particles for the parallel scheme is still very challenging. 


For the estimation of a $2$-dimensional Hamiltonian, we consider the Hamiltonian for a spin-1/2 in a magnetic field, which can be written as
$H(B,\theta,\phi)=B(\sin\theta \cos \phi \sigma_1 +\sin\theta \sin \phi \sigma_2+\cos\theta \sigma_3),$
here $x=(B,\theta, \phi)$ represents the magnitude and the directions of a magnetic field,  $\sigma_1=\left(\begin{array}{cc}
      0 & 1  \\
      1 & 0 \\
          \end{array}\right)$, $\sigma_2=\left(\begin{array}{cc}
      0 & -i  \\
      i & 0 \\
          \end{array}\right)$ and $\sigma_3=\left(\begin{array}{cc}
      1 & 0  \\
      0 & -1 \\
          \end{array}\right)$ are Pauli matrices. The Hamiltonian can also be written concisely as $H(B,\theta,\phi)=B[\vec{n}(\theta,\phi)\cdot \vec{\sigma}]$, where $\vec{n}(\theta,\phi)=(\sin\theta \cos \phi,\sin\theta \sin \phi,\cos\theta)$. We are interested in the ultimate local precision limit in estimating $x=(B, \theta,\phi)$, under the aid of ancillary systems. We first consider the scheme without feedback controls, then extend to the general sequential feedback scheme.

We denote $U(x,T)=e^{-iH(B,\theta,\phi)T}$ as the free evolution of the Hamiltonian with T units of time and $U_A(x,T)=U(x,T)\otimes I_A$ as the evolution with an ancillary system, here $I_A$ denotes the identity operator on the ancillary system. Let $\rho_x=U_A(x,T)\rho_{SA}U_A^\dagger(x,T)$ and $\rho_{x+dx}=U_A(x+dx,T)\rho_{SA}U_A^\dagger(x+dx,T)$, here $\rho_{SA}$ denotes the initial state of system+ancilla and $dx$ represents a small shift of the parameter. The local precision limit of estimating $x$ from the output state $\rho_x$ is related to the Bures distance between $\rho_x$ and $\rho_{x+dx}$ as\cite{HELS67,HOLE82,BRAU94,BRAU96}
 \begin{equation}
\label{eq:BJ}
d^2_{Bures}(\rho_x,\rho_{x+dx})=\sum_{ij}\frac{1}{4}J_{ij}(\rho_x)dx_idx_j,
\end{equation}
here the Bures distance $d_{Bures}$ is defined as
$d_{Bures}(\rho_1,\rho_2)=\sqrt{2-2F(\rho_1,\rho_2)}$
with $F(\rho_1,\rho_2)=\sqrt{\rho_1^{\frac{1}{2}}\rho_2\rho_1^{\frac{1}{2}}}$ as the fidelity between $\rho_1$ and $\rho_2$, and $J_{ij}(\rho_x)$ is the $ij$-th entry of the QFIM $J(\rho_x)$. 
Since
\begin{eqnarray}
\label{eq:maximum}
\aligned
&\max_{\rho_{SA}} d^2_{Bures}(\rho_x,\rho_{x+dx})\\
=&2-2\min_{\rho_{SA}}F(\rho_{SA},U'\otimes I_A\rho_{SA} U'^\dagger\otimes I_A),
\endaligned
\end{eqnarray}
here $U'=U^\dagger(x,T) U(x+dx,T)$, the maximal QFIM is thus related to the minimum fidelity $\min_{\rho_{SA}}F(\rho_{SA},U'\otimes I_A\rho_{SA} U'^\dagger\otimes I_A)$.

For any $d\times d$ unitary $U$, we denote $e^{-iE^{U}_j}$ as the eigenvalues of $U$ with $E^{U}_j\in(-\pi,\pi]$, $1\leq j\leq d$. We call $E^{U}_j$ the eigen-angles of $U$ and assume $E^{U}_{\max}=E^{U}_1\geq E^{U}_2\geq \cdots \geq E^{U}_d=E^{U}_{\min}$ are arranged in decreasing order. It is known that
$\min_{\rho_0}F(\rho_0,U\rho_0 U^\dagger)=\cos\frac{E^{U}_{\max}-E^{U}_{\min}}{2}$ if $E^{U}_{\max}-E^{U}_{\min}\leq \pi$\cite{Fung2}. Denote $C(U)=\frac{E^{U}_{\max}-E^{U}_{\min}}{2}$ then the equation can be written concisely as $\min_{\rho_0}F(\rho_0,U\rho_0 U^\dagger)=\cos C(U)$. Since $E^{U\otimes I_A}_{\max}=E^{U}_{\max}$ and $E^{U\otimes I_A}_{\min}=E^{U}_{\min}$, we also have $\min_{\rho_{SA}}F(\rho_{SA},U\otimes I_A\rho_{SA} U^\dagger\otimes I_A)=\cos C(U)$(we note that this does not mean ancillary system is not useful, the role of ancillary system will be clear later).  

With Eq.(\ref{eq:BJ}) and Eq.(\ref{eq:maximum}) we can then obtain
\begin{eqnarray}
\label{eq:defJ}
\aligned
 \sum_{ij}J_{ij}^{\max}dx_idx_j 
=8\{1-\cos C[U^\dagger(x,T)U(x+dx,T)]\}.
\endaligned
\end{eqnarray}
If $U(x,T)$ is continuous with $x$, then when $dx$ is sufficiently small, $U^\dagger(x,T) U(x+dx,T)\rightarrow I$, $C[U^\dagger(x,T) U(x+dx,T)]\rightarrow 0$, thus up to the second order
\begin{eqnarray}
\label{eq:maxQFIphi2}
\sum_{ij} J_{ij}^{\max}dx_idx_j=4C^2[U^\dagger(x,T) U(x+dx,T)].
  \end{eqnarray}
To ensure there exists a QFIM $J(\rho_x)$ achieves the $J^{\max}$ for all $dx$, we need to show the optimal state $\rho_{SA}$ that achieves the maximum Bures distance in Eq.(\ref{eq:maximum}) is independent of $dx$. In the supplemental material we show that any maximally entangle state(which are those states such that the reduced state is completely mixed, i.e., $Tr_A(\rho_{SA})=1/2I$) achieves the maximum Bures distance in Eq.(\ref{eq:maximum}) for all $dx$, $J^{\max}$ thus corresponds to the QFIM of any maximally entangled probe state. And the maximal QFIM is given by\cite{supp}
\begin{equation}
\label{eq:JT}
 J^{\max}=4\left(\begin{array}{ccc}
      T^2 & 0 & 0  \\
      0 & \sin^2(BT) & 0\\
      0 & 0 & \sin^2(BT)\sin^2(\theta)
          \end{array}\right).
          \end{equation}
Furthermore the projective measurement in the Bell-basis saturates the quantum Cram\'er-Rao bound. In the supplemental material we showed that the distribution of the measurement results in the Bell-basis 
is given by $p_1=\cos^2(BT)$, $p_2 =\sin^2(BT)\cos^2\theta$, $p_3 =\sin^2(BT)\sin^2\theta\cos^2\phi$ and $p_4=\sin^2(BT)\sin^2\theta\sin^2\phi$,
which has the classical Fisher information matrix equals to $J^{\max}$. The quantum Cram\'er-Rao bound is thus saturable and $J^{\max}$ sets the local precision limit when the dynamics is evolved for $T$ units of time. This is consistent with previous studies\cite{Manuel2004,Fujiwara2001}, however our method makes it easy to incorporate feedback controls as we now show.

For the general sequential feedback scheme as in Fig.\ref{fig:scheme}(a), the total evolution can be written as $U_{FA}(x,Nt)=U_NU_A(x,t)\cdots U_2U_A(x,t)U_1 U_A(x,t),$ where $U_A(x,t)=e^{-iH(x)t}\otimes I_A$ with $t=\frac{T}{N}$, and $U_1, U_2,\cdots ,U_N$ denote the feedback controls. 
It can be shown that  $C[U_{FA}^\dagger(x,Nt)U_{FA}(x+dx,Nt)]\leq NC[U_A^\dagger(x,t)U_A(x+dx,t)]$ where the equality can be achieved when $U_1=U_2=\cdots=U_N=U_A^\dagger(x,t)$(see supplementary material\cite{supp}). In practice the true value $x$ is not known a-priori, the estimated value $\hat{x}$ need to be used and the controls $U_1=U_2=\cdots=U_N=U_A^\dagger(\hat{x},t)$ need be updated adaptively. This, however, does not affect the asymptotical scaling\cite{FUJI06, NAGA88,HAYA08}.

From Eq.(\ref{eq:maxQFIphi2}) we then have
\begin{eqnarray}
\label{eq:JN0}
\aligned
\sum_{ij}(J_N^{\max})_{ij}dx_idx_j&= 4C^2[U_{FA}^\dagger(x,Nt) U_{FA}(x+dx, Nt))\\
&\leq 4N^2C^2[U_A^\dagger(x,t)U_A(x+dx,t)]\\
&=N^2 \sum_{ij} (J_1^{\max})_{ij}dx_idx_j,
\endaligned
\end{eqnarray}
thus
\begin{eqnarray}
\label{eq:JN}
\aligned
J^{\max}_N&\leq N^2J^{\max}_1\\
&=4N^2\left(\begin{array}{ccc}
      t^2 & 0 & 0  \\
      0 & \sin^2(Bt) & 0\\
      0 & 0 & \sin^2(Bt)\sin^2(\theta)
          \end{array}\right),
\endaligned
\end{eqnarray}
           here the equality can be saturated asymptotically with the controls $U_1=U_2=\cdots=U_N=U_A^\dagger(\hat{x},t)=e^{iH(\hat{x})t}\otimes I_A$.
         %
In this case the feedback controls only act on the system, we can thus write
\begin{eqnarray}
\aligned
&U_{FA}^\dagger(x,Nt)U_{FA}(x+dx,Nt)=U'\otimes I_A\\
&=e^{ia_{Nt}(x,dx)[\vec{k}_{Nt}(x,dx)\cdot \vec{\sigma}]}\otimes I_A,
\endaligned
\end{eqnarray}
for the last equation we used the fact that any $U'$ can be written as $e^{ia_{Nt}(x,dx)(\vec{k}_{Nt}(x,dx)\cdot \vec{\sigma})}$ where $\vec{k}_{Nt}(x,dx)$ is a unit vector. This has similar form as the free evolution, thus by following the same line of argument one can show the optimal probe state is any maximally entangled state which has the QFIM $J^{\max}_N=4N^2\left(\begin{array}{ccc}
      t^2 & 0 & 0  \\
      0 & \sin^2(Bt) & 0\\
      0 & 0 & \sin^2(Bt)\sin^2(\theta)
          \end{array}\right)$ under the optimal feedback scheme.
In this case the measurement in the Bell basis also saturates the quantum Cram\'er-Rao bound $nCov(\hat{x})\geq (J^{\max}_{N})^{-1}$(see supplementary material for detail), $J^{\max}_N$ thus quantifies the asymptotical precision limit.


To ease comparison with previous results, we rewrite the Hamiltonian as $H=x_1\sigma_1+x_2\sigma_2+x_3\sigma_3$ with $x_1=B\sin\theta\cos\phi$, $x_2=B\sin\theta\sin\phi$, $x_3=B\cos\theta$. In the asymptotical limit the estimation is in the vicinity of the actual value, we can thus write 


\begin{eqnarray}
\nonumber
\aligned
\delta \hat{x}_1&=\sin\theta\cos\phi\delta \hat{B}+B\cos\theta\cos\phi\delta\hat{\theta}-B\sin\theta\sin\phi\delta\hat{\phi},\\
\delta \hat{x}_2&=\sin\theta\sin\phi\delta\hat{B}+B\cos\theta\sin\phi\delta\hat{\theta}+B\sin\theta\cos\phi\delta\hat{\phi},\\
\delta \hat{x}_3&=\cos\theta\delta\hat{B}-B\sin\theta\delta\hat{\theta},\\
\endaligned
\end{eqnarray}
it is then easy to get
$\delta \hat{x_1}^2+\delta \hat{x_2}^2+\delta \hat{x_3}^2=\delta \hat{B}^2+B^2\delta \hat{\theta}^2+B^2\sin^2(\theta)\delta\hat{\phi}^2$.
This will be taken as the figure of merit for comparison as it is used in previous studies\cite{Imai2007,Datta}, which corresponds to take $G=I$ in $Tr[Cov(\hat{x})G]$ under the representation of $(x_1,x_2,x_3)$. We note that the choice of $G=I$ here is just for the purpose of comparison, the precision limit obtained under the feedback scheme is optimal for any $G$---as the obtained precision saturates the quantum Cram\'er-Rao bound $nCov(\hat{x})\geq (J^{\max}_N)^{-1}$, thus for any choices of $G$ it also saturates the lower bound $n Tr[Cov(\hat{x})G]\geq Tr[(J^{\max}_{N})^{-1}G]$.  Here $n$ is the number of times that the procedure is repeated, which accounts for the classical effect, for the following we will neglect $n$ by assuming the procedure is repeated with the same(sufficiently large) number of times.  

We now compare $\delta \hat{x_1}^2+\delta \hat{x_2}^2+\delta \hat{x_3}^2$ obtained from different schemes. Under the optimal sequential feedback scheme we have $Cov(\hat{x})= (J_N^{\max})^{-1}$ with $J^{\max}_N=4N^2\left(\begin{array}{ccc}
      t^2 & 0 & 0  \\
      0 & \sin^2(Bt) & 0\\
      0 & 0 & \sin^2(Bt)\sin^2(\theta)
          \end{array}\right)$, thus
\begin{eqnarray}
\label{eq:scaling}
\aligned
\delta \hat{x_1}^2+\delta \hat{x_2}^2+\delta \hat{x_3}^2&=\delta \hat{B}^2+B^2\delta \hat{\theta}^2+B^2\sin^2(\theta)\delta\hat{\phi}^2\\
&=\frac{1}{4N^2}\left[\frac{1}{t^2}+\frac{2B^2}{\sin^2(Bt)}\right].
\endaligned
\end{eqnarray}
Under the parallel scheme the precision has been extensively studied previously \cite{Peres2001,Acin01,Chiribella2004,Bagan2004,Bagan20042,Chiribella2005,Fujiwara2001,Imai2007,Manuel2004,Manuel20042,Hayashi2006,Datta} with the highest precision given by $Cov(\hat{x})=\frac{3}{N(N+2)}(J_1^{\max})^{-1}$\cite{Imai2007, Datta}, here $J_1^{\max}=4\left(\begin{array}{ccc}
      t^2 & 0 & 0  \\
      0 & \sin^2(Bt) & 0\\
      0 & 0 & \sin^2(Bt)\sin^2(\theta)
          \end{array}\right)$. This corresponds to
\begin{eqnarray}
\label{eq:parallel}
\aligned
\delta \hat{x_1}^2+\delta \hat{x_2}^2+\hat{x_3}^2=\frac{3}{4N(N+2)}\left[\frac{1}{t^2}+\frac{2B^2}{\sin^2(Bt)}\right].
\endaligned
\end{eqnarray}
Compare Eq.(\ref{eq:scaling}) and Eq.(\ref{eq:parallel}) we can see that the optimal sequential feedback scheme has a 3-fold improvement over the optimal parallel scheme.

For a given $T$, when $N\rightarrow \infty$, $t=\frac{T}{N}\rightarrow 0$, $\frac{B^2}{\sin^2(Bt)}\rightarrow \frac{1}{t^2}$, the precision limit under the optimal sequential feedback scheme thus reaches
$\delta \hat{x_1}^2+\delta \hat{x_2}^2+\delta \hat{x_3}^2=\frac{3}{4N^2t^2}=\frac{3}{4T^2}.$
 Note that for the estimation of a single parameter $x_i$ the highest precision one can get within $T$ units of time is $\delta \hat{x_i}^2=\frac{1}{4N^2t^2}=\frac{1}{4T^2}$\cite{GIOV06}. It is conventionally believed that for simultaneous estimation of different parameters of a magnetic field some tradeoffs have to be made on the probe states and measurements, thus not possible to achieve the highest precision for all parameters simultaneously. While the tradeoffs are indeed unavoidable under the parallel scheme, we showed that, contrary to this conventional belief, the optimal sequential feedback scheme can achieve the highest precision for all three parameters of a magnetic field simultaneously.

We next show that for the estimation of general Hamiltonian for $SU(d)$ dynamics the sequential feedback scheme has similar improvement over the parallel scheme.

Given an $SU(d)$ dynamics aided with ancillary system, $U(x,t)=e^{i\sum_{j=1}^{d^2-1}x_jF_jt}\otimes I_A$, here $\{F_j\}$ are traceless self-adjoint matrices and $Tr(F_jF_k)=\delta_{jk}$, i.e., $\{iF_j\}$ form an orthogonal basis of $su(d)$, $x=(x_1,x_2,\cdots,x_{d^2-1})$ are the parameters to be estimated. We compare three schemes: 1)the independent scheme; 2)the parallel scheme; 3)the sequential feedback scheme. The independent scheme is to divide the $N$ uses of the dynamics into $d^2-1$ groups and use $\frac{N}{d^2-1}$ dynamics in each group to estimate one parameter. Under this scheme the variance of each parameter $\delta x_j^2\propto \frac{1}{(\frac{N}{d^2-1})^2t^2}=\frac{(d^2-1)^2}{N^2t^2}$, the summation of variance is then $\sum_{j=1}^{d^2-1}\delta x_j^2\propto \frac{(d^2-1)^3}{N^2t^2}$. For the parallel scheme the minimum summation of variance has been obtained previously as $\sum_{j=1}^{d^2-1}\delta x_j^2=\frac{d(d+1)(d^2-1)}{4N(N+d)t^2}$\cite{Imai2007}.
For the sequential feedback scheme, we show that(see supplemental material\cite{supp}) by taking the maximally entangled state as the probe state and using the optimal feedback control $U_1=U_2=\cdots=U_N=U^\dagger_A(\hat{x},t)$, the quantum Fisher information matrix is given by $\frac{4N^2t^2}{d}I$, and the quantum Cram\'er-Rao bound can be saturated. The summation of variance under the optimal feedback scheme is thus given by $\sum_{j=1}^{d^2-1}\delta x_j^2=\frac{d(d^2-1)}{4N^2t^2}$, which has an order of $O(d+1)$ improvement over the parallel scheme and an order of $O(d^3)$ improvement over the independent scheme.

\emph{Discussion and conclusion:}
The comparison between the sequential feedback scheme and the parallel scheme has been a subject of lasting interest in quantum channel discrimination and quantum parameter estimation. In quantum channel discrimination Ac$\acute{i}$n\cite{Acin01} and \citet{Mauro2001} studied the optimal parallel scheme for the discrimination between two unitary dynamics, \citet{Duan2007} then showed the sequential feedback scheme is equivalent to the parallel scheme for the discrimination of unitary dynamics, \citet{ChiribellaDP08} showed the sequential feedback scheme can outperform the parallel scheme for discriminating quantum channels with memory effects. Optimal sequential scheme has also been obtained for the discrimination of two general quantum channels\cite{DuanFY09}. For single-parameter quantum estimation the sequential feedback scheme is shown to be equivalent to the parallel scheme under unitary\cite{GIOV06} and dephasing dynamics\cite{Boxio2012,Rafal2014}, and it has been conjectured that asymptotically the sequential feedback scheme is equivalent to the parallel scheme\cite{Rafal2014}. For multi-parameter quantum estimation \citet{humphreys2013} showed the parallel scheme has an order of $O(d)$ improvement over the independent scheme for estimating $d$ parameters with commutating generators, for general unitary dynamics the optimal parallel scheme has also been studied\cite{Imai2007,Datta} which shows the parallel scheme has similar improvement over the independent scheme. 

Prior of this study a general belief has been that under unitary dynamics the sequential feedback scheme is equivalent to the parallel scheme(while under noisy dynamics the sequential feedback scheme is believed to be either equivalent to the parallel scheme or can only outperform the parallel scheme for channels with special properties). Here by showing the sequential feedback scheme has an order of $O(d+1)$ improvement over the parallel scheme for the estimation of $SU(d)$ dynamics, our study disclosed a unique feature for multi-parameter quantum estimation and deepened the understanding on the relationship between the sequential feedback scheme and the parallel scheme.


Our study also sets a benchmark on the local precision limit for the estimation of a magnetic field, which is of practical importance for many applications. The precision is obtained by optimizing all steps in the procedure of the estimation, thus represents the ultimate precision one can achieve for the estimation of a magnetic field asymptotically. Our study shows that it is possible to achieve the highest precision simultaneously for all three parameters of a magnetic field, contrary to the conventional belief that some tradeoffs have to be made on the precision of different parameters. This opened the possibility and initiated the study of using feedback controls to counteract the tradeoffs in multi-parameter quantum estimation. 
Future research includes finding the ultimate precision at the presence of general noises.

In the supplemental material we also discussed the possible extension to Hamiltonian parameter estimation with a prior distribution, and showed that the feedback scheme gains over the parallel scheme through adaptive choice of the evolution time\cite{supp}. Intuitively the feedback scheme gains over the parallel scheme by utilizing the information encoded in the prior distribution to design the feedback controls, while under the parallel scheme the information is ignored during the evolution stage. Future research includes quantifying the gain of the feedback scheme exactly under any prior distribution.

\appendix
\section{Supplemental Material}
\subsection{Optimal probe states}
We show that the maximally entangled states are the optimal probe states that achieve $J^{\max}$ in the estimation of a magnetic field.

Consider the dynamics $U(x,T)=e^{-iH(B,\theta,\phi)T}$, where $H(B,\theta,\phi)=B(\sin\theta \cos \phi \sigma_1 +\sin\theta \sin \phi \sigma_2+\cos\theta \sigma_3),$ for any pure state $|\varphi_{SA}\rangle$, let $\rho_x=U(x,T)\otimes I_A |\varphi_{SA}\rangle \langle \varphi_{SA}|U^\dagger(x,T)\otimes I_A$, $\rho_{x+dx}=U(x+dx,T)\otimes I_A |\varphi_{SA}\rangle \langle \varphi_{SA}|U^\dagger(x+dx,T)\otimes I_A$. As shown in the main text, the optimal probe states are those states that minimize $F(\rho_x,\rho_{x+dx})$ for all $dx$, note that
\begin{eqnarray}
\label{eq:optimalstate}
\aligned
&F(\rho_x,\rho_{x+dx})\\
=&F[U(x,T)\otimes I_A |\varphi_{SA}\rangle \langle \varphi_{SA}|U^\dagger(x,T)\otimes I_A, \\
&U(x+dx,T)\otimes I_A |\varphi_{SA}\rangle \langle \varphi_{SA}|U^\dagger(x+dx,T)\otimes I_A]\\
=&F(|\varphi_{SA}\rangle \langle \varphi_{SA}|, U'\otimes I_A|\varphi_{SA}\rangle \langle \varphi_{SA}|U'^\dagger\otimes I_A)\\
=&|\langle \varphi_{SA}| U'\otimes I_A|\varphi_{SA}\rangle|\\
=&|Tr(\rho_SU')|,
\endaligned
\end{eqnarray}
where $\rho_S=Tr_A(|\varphi_{SA}\rangle \langle \varphi_{SA}|)$ and $U'=U^\dagger(x,T) U(x+dx,T)$ which can be written as $U'=e^{ia(x,dx)[\hat{k}(x,dx)\cdot \hat{\sigma}]}$ with $\hat{k}(x,dx)$ as a unit vector.
The eigenvalues of $e^{ia(x,dx)[\hat{k}(x,dx)\cdot \hat{\sigma}]}$ are $e^{\pm i a(x,dx)}$, we can thus diagonalize $U'$ as
$\tilde{U}(\hat{k})\left(\begin{array}{cc}
      e^{ia(x,dx)} & 0 \\
      0 & e^{-ia(x,dx)} \\
                \end{array}\right)\tilde{U}^\dagger(\hat{k})$
where $\tilde{U}(\hat{k})$ is a unitary which depends on $\hat{k}(x,dx)$. Let $\tilde{\rho}=\tilde{U}^\dagger(\hat{k}) \rho_S \tilde{U}(\hat{k})$, then
\begin{eqnarray}
\aligned
|Tr(\rho_SU')|&=|Tr[\tilde{\rho}\left(\begin{array}{cc}
      e^{ia} & 0 \\
      0 & e^{-ia} \\
                \end{array}\right)]|\\
                &=|\tilde{\rho}_{11}e^{ia}+\tilde{\rho}_{22}e^{-ia}|\\
&=\sqrt{\cos^2(a)+(\tilde{\rho}_{11}-\tilde{\rho}_{22})^2\sin^2(a)}\\
&\geq |\cos(a)|,
\endaligned
\end{eqnarray}
the minimum is achieved when $\tilde{\rho}_{11}=\tilde{\rho}_{22}=\frac{1}{2}$, i.e., when $\tilde{\rho}=\frac{1}{2}I$, from which we obtain the optimal $\rho_S=\tilde{U}(\hat{k})\tilde{\rho} \tilde{U}^\dagger(\hat{k})=\frac{1}{2}I$(note that other choices of $\tilde{\rho}$ will not be invariant under the conjugation of $\tilde{U}(\hat{k})$, which then does not lead to any fixed probe state). The optimal probe state $|\varphi_{SA}\rangle$ is thus any maximally entangled state.

\subsection{The maximal quantum Fisher information matrix for the estimation of a magnetic field under free evolution}
In this section we show how to calculate $J^{\max}$ in the estimation of a magnetic field under free evolution.
In the main text we showed that
\begin{eqnarray}
\label{eq:defJsm}
\aligned
 \sum_{ij}J_{ij}^{\max}dx_idx_j 
=8\{1-\cos C[U^\dagger(x,T) U(x+dx,T)]\}.
\endaligned
\end{eqnarray}
To calculate $J^{\max}$, let
\begin{eqnarray}
\label{eq:U'}
\aligned
U'&=U^\dagger(x,T) U(x+dx,T)\\
&=e^{i BT[\vec{n}(\theta,\phi)\cdot \vec{\sigma}]}e^{-i (B+dB)T[\vec{n}(\theta+d\theta,\phi+d\phi)\cdot \vec{\sigma}]}\\
&=e^{ia(x,dx)[\vec{k}(x,dx)\cdot \vec{\sigma}]}
\endaligned
\end{eqnarray}
where $\vec{k}(x,dx)$ is a unit vector, $a(x,dx)\geq 0$ which can be easily obtained from
$\cos[a(x,dx)]=\cos(BT)\cos[(B+dB)T]+\vec{n}(\theta,\phi)\cdot\vec{n}(\theta+d\theta,\phi+d\phi) \sin(BT)\sin[(B+dB)T]$. Since the eigenvalues of $e^{ia(\vec{k}\cdot \vec{\sigma})}$ are $e^{\pm ia}$ we have $E^{U'}_{\max}=a(x,dx)$, $E^{U'}_{\min}=-a(x,dx)$, thus $C[U^\dagger(x,T) U(x+dx,T)]=\frac{E^{U'}_{\max}-E^{U'}_{\min}}{2}=a(x,dx)$.
Using Eq.(\ref{eq:U'}) we can expand the right side of Eq.(\ref{eq:defJsm}) to the second order
\begin{eqnarray}
\aligned
&8\{1-\cos C[U^\dagger(x,T) U(x+dx,T)]\}\\
=&8[1-\cos a(x,dx)]\\
=&4[T^2dB^2+\sin^2(BT)d\theta^2+\sin^2\theta\sin^2(BT)d\phi^2],
\endaligned
\end{eqnarray}
the maximum QFIM is then obtained by comparing coefficients at both sides of Eq.(\ref{eq:defJsm}), which gives
\begin{equation}
\label{eq:JTsm}
 J^{\max}=4\left(\begin{array}{ccc}
      T^2 & 0 & 0  \\
      0 & \sin^2(BT) & 0\\
      0 & 0 & \sin^2(BT)\sin^2(\theta)
          \end{array}\right).
          \end{equation}

We further show that this $J^{\max}$ dominates all other QFIMs under free evolution, i.e., $J^{\max} \geq J(\rho_x)$ for all other $J(\rho_x)$($A\geq B$ means $A-B$ is positive semi-definite). To see this take any probe state $\rho_{SA}$, let $\rho_x=U(x,T)\otimes I_A(\rho_{SA})$, $\rho_{x+dx}=U(x+dx,T)\otimes I_A(\rho_{SA})$, here $U(x+dx,T)\otimes I_A(\rho_{SA})=U(x+dx,T)\otimes I_A \rho_{SA}U^\dagger(x+dx,T)\otimes I_A$, then for any $dx=(dB,d\theta,d\phi)$(which is always assumed to be sufficiently small so the second order expansion is valid), we have
\begin{eqnarray}
\aligned
&\sum_{ij}\frac{1}{4}J_{ij}(\rho_x)dx_idx_j\\
&=d^2_{Bures}(\rho_x,\rho_{x+dx})\\
&=d^2_{Bures}[U(x,T)\otimes I_A(\rho_{SA}),U(x+dx,T)\otimes I_A(\rho_{SA})]\\
&\leq \max_{\rho_{SA}}d^2_{Bures}[U(x,T)\otimes I_A(\rho_{SA}),U(x+dx,T)\otimes I_A(\rho_{SA})]\\
&=\sum_{ij}\frac{1}{4}J_{ij}^{\max}dx_idx_j,
\endaligned
\end{eqnarray}
where the last equality is achieved by taking the $\rho_{SA}$ as the maximally entangled state. Thus $dx J^{\max} dx^T\geq dx J(\rho_x)dx^T$ for all $dx$, i.e., $J^{\max}\geq J(\rho_x)$, $J^{\max}$ thus sets the precision limit for estimating $x$. This is consistent with previous studies\cite{Manuel2004,Fujiwara2001}, however our method makes it easy to incorporate feedback controls.

\subsection{Measurement in the Bell basis saturates the quantum Cram\'er-Rao bound}
In this section we show that there exists POVM saturates the quantum Cram\'er-Rao bound in estimating a magnetic field under the optimal sequential feedback scheme, one can also see Sec.\ref{sec:sud} of this supplementary material for an alternative proof.

We first show that under the free evolution $U_A(x,T)=e^{-iH(B,\theta,\phi)T}\otimes I_A$ the measurement in the Bell basis saturates the quantum Cram\'er-Rao bound, then extend to the feedback scheme.

With the maximally entangled state $|\varphi_{\max}\rangle= \frac{1}{\sqrt{2}}(\begin{bmatrix}
0 \\
1
\end{bmatrix}\otimes \begin{bmatrix}
0\\
1
\end{bmatrix}+\begin{bmatrix}
1 \\
0
\end{bmatrix}\otimes \begin{bmatrix}
1\\
0
\end{bmatrix})$ as the probe state, the output state under the free evolution $e^{-iH(B,\theta,\phi)T}\otimes I_A$ is
\begin{eqnarray}
\aligned
|\varphi(B,\theta,\phi)\rangle &=\frac{1}{\sqrt{2}}\begin{bmatrix}
i\sin(BT)\sin\theta e^{-i\phi} \\
\cos(BT)-i\sin(BT)\cos\theta
\end{bmatrix}\otimes \begin{bmatrix}
0\\
1
\end{bmatrix}\\
&+\frac{1}{\sqrt{2}}\begin{bmatrix}
\cos(BT)+i\sin(BT)\cos\theta\\
i\sin(BT)\sin\theta e^{i\phi}
\end{bmatrix}\otimes \begin{bmatrix}
1\\
0
\end{bmatrix},
\endaligned
\end{eqnarray}
which can be rewritten as
\begin{eqnarray}
\aligned
|\varphi(B,\theta,\phi)\rangle =&\frac{\cos(BT)}{\sqrt{2}}(|00\rangle+|11\rangle)\\
&-\frac{i\sin(BT)\cos\theta}{\sqrt{2}}(|00\rangle-|11\rangle)\\
&+\frac{i\sin(BT)\sin\theta\cos\phi}{\sqrt{2}}(|10\rangle+|01\rangle)\\
&+\frac{\sin(BT)\sin\theta\sin\phi}{\sqrt{2}}(|10\rangle-|01\rangle).
\endaligned
\end{eqnarray}
The probability of the measurement results in the Bell-basis
\begin{eqnarray}
\aligned
|\varphi_1\rangle &=\frac{|00\rangle+|11\rangle}{\sqrt{2}},\\
|\varphi_2\rangle &=\frac{|00\rangle-|11\rangle}{\sqrt{2}},\\
|\varphi_3\rangle &=\frac{|10\rangle+|01\rangle}{\sqrt{2}},\\
|\varphi_4\rangle &=\frac{|10\rangle-|01\rangle}{\sqrt{2}}\\
\endaligned
\end{eqnarray}
can then be easily obtained as
\begin{eqnarray}
\aligned
p_1 &=\cos^2(BT),\\
p_2 &=\sin^2(BT)\cos^2\theta,\\
p_3 &=\sin^2(BT)\sin^2\theta\cos^2\phi,\\
p_4 &=\sin^2(BT)\sin^2\theta\sin^2\phi.\\
\endaligned
\end{eqnarray}
Based on the probability distribution it is straightforward to calculate the classical Fisher information matrix as
 $4\left(\begin{array}{ccc}
      T^2 & 0 & 0  \\
      0 & \sin^2(BT) & 0\\
      0 & 0 & \sin^2(BT)\sin^2(\theta)
          \end{array}\right),$
which is the same as the quantum Fisher information matrix. The quantum Cram\'er-Rao bound is thus asymptotically saturable in this case. This is consistent with previous studies\cite{Manuel2004}.

Under the optimal feedback scheme the total dynamics is given by $U_{FA}(x,Nt)=U_NU_A(x,t)\cdots U_2U_A(x,t)U_1 U_A(x,t),$ where $U_A(x,t)=e^{-iH(x)t}\otimes I_A$ and the feedback controls $U_1=U_2=\cdots=U_A^\dagger(\hat{x},t)$. Since both the system dynamics and the feedback controls only act on the system, the total dynamics can be written as  $U_{FA}(x,Nt)=e^{-i\tilde{B}[\sin\tilde{\theta}\cos\tilde{\phi}\sigma_x+\sin\tilde{\theta}\sin\tilde{\phi}\sigma_y+\cos\tilde{\theta}\sigma_z]}\otimes I_A$ where $\tilde{B}(B,\theta,\phi),\tilde{\theta}(B,\theta,\phi),\tilde{\phi}(B,\theta,\phi)$ are some functions of $x=(B,\theta, \phi)$. With the maximally entangled state
$|\varphi_{\max}\rangle= \frac{1}{\sqrt{2}}(\begin{bmatrix}
0 \\
1
\end{bmatrix}\otimes \begin{bmatrix}
0\\
1
\end{bmatrix}+\begin{bmatrix}
1 \\
0
\end{bmatrix}\otimes \begin{bmatrix}
1\\
0
\end{bmatrix})$ as the probe state, the output state is similarly given by
\begin{eqnarray}
\aligned
|\varphi(B,\theta,\phi)\rangle &=\frac{1}{\sqrt{2}}\begin{bmatrix}
i\sin(\tilde{B}T)\sin\tilde{\theta}e^{-i\tilde{\phi}} \\
\cos(\tilde{B}T)-i\sin(\tilde{B}T)\cos\tilde{\theta}
\end{bmatrix}\otimes \begin{bmatrix}
0\\
1
\end{bmatrix}\\
&+\frac{1}{\sqrt{2}}\begin{bmatrix}
\cos(\tilde{B}T)+i\sin(\tilde{B}T)\cos\tilde{\theta}\\
i\sin(\tilde{B}T)\sin\tilde{\theta}e^{i\tilde{\phi}}
\end{bmatrix}\otimes \begin{bmatrix}
1\\
0
\end{bmatrix}.
\endaligned
\end{eqnarray}
It is intuitively clear that better precision of $(\tilde{B},\tilde{\theta},\tilde{\phi})$ corresponds to better precision of $(B,\theta,\phi)$, the measurement in the Bell basis thus is also optimal. To see this more formally, let $\rho_x=|\varphi(B,\theta,\phi)\rangle\langle\varphi(B,\theta,\phi)|$ and denote $L_{B},L_{\theta},L_{\phi}$ as the symmetrical logarithm derivatives of $B,\theta,\phi$ respectively, which are the solutions to the following equations
\begin{eqnarray}
\aligned
\frac{\partial \rho_x}{\partial B}&=\frac{1}{2}(\rho_xL_{B}+L_{B}\rho_x),\\
\frac{\partial \rho_x}{\partial \theta}&=\frac{1}{2}(\rho_xL_{\theta}+L_{\theta}\rho_x),\\
\frac{\partial \rho_x}{\partial \phi}&=\frac{1}{2}(\rho_xL_{\phi}+L_{\phi}\rho_x).\\
\endaligned
\end{eqnarray}
To show the quantum Cram\'er-Rao bound is achievable we just need to show $Im[\langle\varphi(B,\theta,\phi)|L_iL_j|\varphi(B,\theta,\phi)\rangle]=0$ $\forall L_i,L_j\in \{L_{B},L_{\theta},L_{\phi}\}$\cite{Matsumoto2002,Fujiwara2001}.

We first denote $L_{\tilde{B}},L_{\tilde{\theta}},L_{\tilde{\phi}}$ as the symmetrical logarithm derivatives of $\tilde{B},\tilde{\theta},\tilde{\phi}$ respectively, which are solutions to the following equations
\begin{eqnarray}
\aligned
\frac{\partial \rho_x}{\partial \tilde{B}}&=\frac{1}{2}(\rho_xL_{\tilde{B}}+L_{\tilde{B}}\rho_x),\\
\frac{\partial \rho_x}{\partial \tilde{\theta}}&=\frac{1}{2}(\rho_xL_{\tilde{\theta}}+L_{\tilde{\theta}}\rho_x),\\
\frac{\partial \rho_x}{\partial \tilde{\phi}}&=\frac{1}{2}(\rho_xL_{\tilde{\phi}}+L_{\tilde{\phi}}\rho_x).\\
\endaligned
\end{eqnarray}
 It is known that $Im[\langle\varphi(B,\theta,\phi)|L_iL_j|\varphi(B,\theta,\phi)\rangle]=0$ $\forall L_i,L_i\in\{L_{\tilde{B}},L_{\tilde{\theta}},L_{\tilde{\phi}}\}$\cite{Fujiwara2001,Manuel2004} (we also have shown specifically that the measurement in the Bell basis saturates the quantum Cram\'er-Rao bound for $(\tilde{B},\tilde{\theta},\tilde{\phi})$).
Now since
\begin{eqnarray}
\aligned
\frac{\partial \rho_x}{\partial B}&=\frac{\partial \rho_x}{\partial \tilde{B}}\frac{\partial \tilde{B}}{\partial B}+\frac{\partial \rho_x}{\partial \tilde{\theta}}\frac{\partial \tilde{\theta}}{\partial B}+\frac{\partial \rho_x}{\partial \tilde{\phi}}\frac{\partial \tilde{\phi}}{\partial B},\\
\frac{\partial \rho_x}{\partial \theta}&=\frac{\partial \rho_x}{\partial \tilde{B}}\frac{\partial \tilde{B}}{\partial \theta}+\frac{\partial \rho_x}{\partial \tilde{\theta}}\frac{\partial \tilde{\theta}}{\partial \theta}+\frac{\partial \rho_x}{\partial \tilde{\phi}}\frac{\partial \tilde{\phi}}{\partial \theta},\\
\frac{\partial \rho_x}{\partial \phi}&=\frac{\partial \rho_x}{\partial \tilde{B}}\frac{\partial \tilde{B}}{\partial \phi}+\frac{\partial \rho_x}{\partial \tilde{\theta}}\frac{\partial \tilde{\theta}}{\partial \phi}+\frac{\partial \rho_x}{\partial \tilde{\phi}}\frac{\partial \tilde{\phi}}{\partial \phi},\\
\endaligned
\end{eqnarray}
we get
\begin{eqnarray}
\aligned
L_{B}&=\frac{\partial \tilde{B}}{\partial B}L_{\tilde{B}}+\frac{\partial \tilde{\theta}}{\partial B}L_{\tilde{\theta}}+\frac{\partial \tilde{\phi}}{\partial B}L_{\tilde{\phi}},\\
L_{\theta}&=\frac{\partial \tilde{B}}{\partial \theta}L_{\tilde{B}}+\frac{\partial \tilde{\theta}}{\partial \theta}L_{\tilde{\theta}}+\frac{\partial \tilde{\phi}}{\partial \theta}L_{\tilde{\phi}},\\
L_{\phi}&=\frac{\partial \tilde{B}}{\partial \phi}L_{\tilde{B}}+\frac{\partial \tilde{\theta}}{\partial \phi}L_{\tilde{\theta}}+\frac{\partial \tilde{\phi}}{\partial \phi}L_{\tilde{\phi}},\\
\endaligned
\end{eqnarray}
which are just linear combinations of $L_{\tilde{B}},L_{\tilde{\theta}},L_{\tilde{\phi}}$. Thus $Im[\langle\varphi(B,\theta,\phi)|L_iL_j|\varphi(B,\theta,\phi)\rangle]=0$ $\forall L_i,L_j\in \{L_{B},L_{\theta},L_{\phi}\}$, the quantum Cram\'er-Rao bound is then saturable. And as $L_{B},L_{\theta},L_{\phi}$ are just linear combinations of $L_{\tilde{B}},L_{\tilde{\theta}},L_{\tilde{\phi}}$, the measurements that saturate the quantum Cram\'er-Rao bound for $(B,\theta,\phi)$ is then the same as the measurements saturate the bound for $(\tilde{B},\tilde{\theta},\tilde{\phi})$, in particular the measurement in the Bell basis saturates the quantum Cram\'er-Rao bound. This can also be seen in Sec.\ref{sec:sud} of this supplementary material, where we show that the feedback controls essentially shift the parameter values, since the measurement in the Bell basis is optimal for all values of the parameters, this measurement is then also optimal under the feedback scheme.    

\subsection{Optimal feedback control}
For completeness we include a derivation of the optimal feedback control which follows the treatment in \cite{YuanTime}. We derive the case of $N=2$, same strategy works in the general case. When $N=2$, $U_{FA}(x,2t)=U_2U_A(x,t)U_1U_A(x,t)$, here $U_A(x,t)=e^{-iH(B,\theta,\phi)t}\otimes I_A$ and $U_1,U_2$ are feedback controls which can act on system+ancilla. Now
\begin{eqnarray}
\nonumber
\label{eq:align}
\aligned
&U_{FA}^\dagger(x,2t) U_{FA}(x+dx,2t)\\
=& U^\dagger_A(x,t)U_1^\dagger U^\dagger_A(x,t)U_2^\dagger U_2 U_A(x+dx,t)U_1U_A(x+dx,t)\\
=& U^\dagger_A(x,t)U_1^\dagger [U^\dagger_A(x,t) U_A(x+dx,t)]\times\\
&U_1[U_A(x,t)U^\dagger_A(x,t)]U_A(x+dx,t)\\
=&(U^\dagger_A(x,t)U_1^\dagger) [U^\dagger_A(x,t) U_A(x+dx,t)](U_1U_A(x,t))\\
&[U^\dagger_A(x,t)U_A(x+dx,t)].\\
\endaligned
\end{eqnarray}
$U_2$ can be chosen as any unitary since it does not change $C[U_{FA}^\dagger(x,2t) U_{FA}(x+dx,2t)]$. Now divide $U_{FA}^\dagger(x,2t) U_{FA}(x+dx,2t)$ into two parts, $(U^\dagger_A(x,t)U_1^\dagger) [U^\dagger_A(x,t) U_A(x+dx,t)](U_1U_A(x,t))$ and $U^\dagger_A(x,t) U_A(x+dx,t)$, then
\begin{eqnarray}
\label{eq:inq}
\aligned
C&[U_{FA}^\dagger(x,t) U_{FA}(x+dx,2t)]\leq C[U^\dagger_A(x,t) U_A(x+dx,t)]\\
&+C[(U^\dagger_A(x,t)U_1^\dagger) [U^\dagger_A(x,t) U_A(x+dx,t)](U_1U_A(x,t))]\\
&=2C[U^\dagger_A(x,t) U_A(x+dx,t)],
\endaligned
\end{eqnarray}
the first inequality we used the fact $C(U_1U_2)\leq C(U_1)+C(U_2)$ if $C(U_1)+C(U_2)\leq \frac{\pi}{2}$\cite{ChildsPR00,Acin01, Chau2011,YuanTime}; the second equality is based on the fact that $(U^\dagger_A(x,t)U_1^\dagger) [U^\dagger_A(x,t) U_A(x+dx,t)](U_1U_A(x,t))$ has the same eigen-angles as $U^\dagger_A(x,t) U_A(x+dx,t)$.
One choice of control that saturates the equality is let $U_1=U^\dagger_A(x,t)$, as it aligns the eigenvalues of the two parts and the corresponding maximal and minimal eigen-angles add up, in this case $U_{FA}^\dagger(x,2t) U_{FA}(x+dx,2t)=[U^\dagger_A(x,t) U_A(x+dx,t)]^2$, $C[U_{FA}^\dagger(x,2t) U_{FA}(x+dx,2t)]=2C[U_A^\dagger(x,t) U_A(x+dx,t)]$ saturates the equality. This argument can be easily generalized to the general $N$ with  $C[U_{FA}^\dagger(x,Nt) U_{FA}(x+dx,Nt)]\leq NC[U_{A}^\dagger(x,t) U_{A}(x+dx,t)]$ and the equality can be saturated with the controls $U_1=U_2=\cdots=U_{N-1}=U^\dagger_A(x,t)$ and arbitrary $U_N$, which for simplicity, can also be taken as $U_A^\dagger(x,t)$.

\subsection{Estimation of the Hamiltonian for $SU(d)$ dynamics}
\label{sec:sud}
We consider the estimation of general Hamiltonians for $SU(d)$ dynamics under the aid of ancillary systems and feedback controls, with the probe state taken as the maximally entangled state $|\varphi\rangle=\frac{1}{\sqrt{d}}\sum_{p=1}^d|pp\rangle$, where $\{|p\rangle|p=1,2,\cdots, d\}$ form an orthonormal basis of a d-dimensional Hilbert space.
Let $U_A(x,t)=e^{i\sum_{j=1}^{d^2-1}x_jF_jt}\otimes I_A$, here $\{iF_j\}$ form a basis of $su(d)$ and $F_j$ are traceless self-adjoint matrices and chosen as $Tr(F_jF_k)=\delta_{jk}$, $x=(x_1,x_2,\cdots,x_{d^2-1})$ are the interested parameters, $t=\frac{T}{N}$. And $U_{FA}(x,Nt)=U_NU_A(x,t)\cdots U_1U_A(x,t)$, with the controls taking as $U_1=U_2=\cdots U_N=U^\dagger_A(\hat{x},t)$.

When $N\rightarrow \infty$, $t=\frac{T}{N}\rightarrow 0$, then $U_A(x,t)=e^{i\sum_{j=1}^{d^2-1}x_jF_jt}\otimes I_A\approx (I+i\sum_{j=1}^{d^2-1}x_jF_jt)\otimes I_A$, we then have
\begin{eqnarray}
\aligned
&U_iU_A(x,t)\\
=&U_A^\dagger(\hat{x},t)U_A(x,t)\\
\approx&[(I-i\sum_{j=1}^{d^2-1}\hat{x}_jF_jt)\otimes I_A][(I+i\sum_{j=1}^{d^2-1}x_jF_jt)\otimes I_A]\\
\approx&[I+ i\sum_{j=1}^{d^2-1}(x_j-\hat{x}_j)F_jt]\otimes I_A\\
\approx &e^{i\sum_{j=1}^{d^2-1}(x_j-\hat{x}_j)F_jt}\otimes I_A,
\endaligned
\end{eqnarray}
the total dynamics is then
\begin{eqnarray}
\aligned
U_{FA}(x,Nt)&=[U_A^\dagger(\hat{x},t)U_A(x,t)]^N\\
&\approx e^{iN\sum_{j=1}^{d^2-1}(x_j-\hat{x}_j)F_jt}\otimes I_A\\
&=e^{i\sum_{j=1}^{d^2-1}(x_j-\hat{x}_j)F_jT}\otimes I_A.
\endaligned
\end{eqnarray}
In the asymptotical limit $\hat{x}\rightarrow x$, the controls essentially reduce the problem of estimating a general $x$ to estimating $x=(0,0,\cdots, 0)$. In the remain part we will then just consider the precision under the dynamics $U_A(x,T)=e^{i\sum_{j=1}^{d^2-1}x_jF_jT}\otimes I_A$ at the point of $x=(0,0,\cdots,0)$. We follow the treatment in \cite{Manuel2004} to calculate the quantum Fisher information matrix of this dynamics.

The output state in this case is unchanged, $U_A(x,T)|\varphi\rangle=\frac{1}{\sqrt{d}}|pp\rangle$. Denote $L_j$ as the symmetric logarithm derivative of the output state with respect to $x_j$, and $|l_j\rangle=L_j|\varphi\rangle$, then it is known that $|l_j\rangle=2(|\varphi,_j\rangle+\langle \varphi,_j|\varphi\rangle|\varphi\rangle)$ with $|\varphi,_j\rangle=\frac{\partial U_A(x,T)|\varphi\rangle}{\partial x_j}$\cite{Fujiwara2001,Matsumoto2002}. To calculate $|\varphi,_j\rangle$ we can expand $U_A(x,T)\approx (I+i\sum_{j=1}^{d^2-1}x_jF_jT)\otimes I_A$, thus
\begin{eqnarray}
\aligned
&\frac{\partial U_A(x,T)|\varphi\rangle}{\partial x_j}\\
=&\frac{\partial [(I+i\sum_{j=1}^{d^2-1}x_jF_jT)\otimes I_A |\varphi\rangle]}{\partial x_j}|_{x=(0,0,\cdots,0)}\\
=&F_j\otimes I_A|\varphi\rangle T.
\endaligned
\end{eqnarray}
Then
\begin{eqnarray}
\aligned
\langle \varphi,_j|\varphi\rangle&=\langle \varphi |F_j\otimes I_A |\varphi\rangle T\\
&=Tr(F_j\frac{1}{d}I)T\\
&=0,
\endaligned
\end{eqnarray}
where for the second equality we used the fact the reduced state of a maximally entangled state is $\frac{1}{d}I$, for the third equality we used the fact that $F_j$ is traceless. Thus
\begin{equation}
|l_j\rangle=2|\varphi,_j\rangle=2F_j\otimes I_A|\varphi\rangle T,
\end{equation}
it is then easy to see that
\begin{eqnarray}
\aligned
\langle l_k|l_j\rangle&=4T^2\langle \varphi |F_kF_j\otimes I_A |\varphi\rangle\\
&=4T^2Tr(F_kF_j\frac{1}{d}I)\\
&=\frac{4T^2}{d}\delta_{kj}.
\endaligned
\end{eqnarray}
The entries of the quantum Fisher information matrix is given by $J_{kj}=Re(\langle l_k|l_j\rangle)$, we then have $J=\frac{4T^2}{d}I=\frac{4N^2t^2}{d}I$. Since in this case $Im(\langle l_k|l_j\rangle)=0$ the quantum Cram\'er-Rao bound $Cov(\hat{x})\geq J^{-1}$ is thus achievable\cite{Matsumoto2002,Fujiwara2001}, we then have $\sum_{j=1}^{d^2-1}\delta x_j^2=\frac{d(d^2-1)}{4N^2t^2}$.

One measurement that saturates the quantum Cram\'er-Rao bound in this case can be taken as the projective measurements on the basis consisting with $(F_i\otimes I_A)\sum_{p=1}^{d}|pp\rangle$, $i=1,\cdots, d^2-1$, and $\frac{1}{\sqrt{d}}\sum_{p=1}^d|pp\rangle$. As when $(x_1,x_2,\cdots,x_{d^2-1})\rightarrow (0,0,\cdots,0)$, the final state can be approximated as  $[(I+i\sum_{j=1}^{d^2-1}x_jF_jT)\otimes I_A]\frac{1}{\sqrt{d}}\sum_{p=1}^d|pp\rangle$, the probability
under the chosen projective measurement can be calculated as $p_1=\frac{x_1^2T^2}{d}, p_2=\frac{x_2^2T^2}{d}, \cdots, p_{d^2-1}=\frac{x_{d^2-1}^2T^2}{d}$ and
$p_{0}=1-\sum_{i=1}^{d^2-1}p_i$. It is then straightforward to verify that when $(x_1,x_2,\cdots,x_{d^2-1})\rightarrow (0,0,\cdots,0)$ the classical Fisher information matrix is $\frac{4T^2}{d}I$, which is the same as the quantum Fisher information matrix. It is also easy to see that when $d=2$, this measurement can be reduced to the projective measurements in the Bell basis by choosing $F_i$ as Pauli matrices.

The quantum Fisher information matrix under the optimal parallel scheme is given by $J_N=\frac{N(N+d)}{d+1}J_1$\cite{Imai2007} while in this case $J_1=\frac{4t^2}{d}I$, thus $\sum_{j=1}^{d^2-1}\delta x_j^2=\frac{(d+1)d(d^2-1)}{4N(N+d)t^2}$ which is $O(d+1)$-times bigger than the value obtained under the feedback scheme.

\subsection{The gain of the optimal feedback scheme over the optimal parallel scheme for the estimation of a magnetic field}
 In this section we give some further analysis on the gain of the feedback scheme with a prior distribution, for which we will focus on the the estimation of a magnetic field.


Under the total dynamics $U_{FA}(x,Nt)=U_NU_A(x,t)\cdots U_1U_A(x,t)$, with the controls taking as $U_1=U_2=\cdots U_N=U^\dagger_A(\hat{x}_C,t)$(here $\hat{x}_C$ denote the estimated value used in the control), when $N$ is sufficiently large the feedback controls shift the parameter from the true value $x=B(\sin\theta\cos\phi,\sin\theta\sin\phi,\cos\theta)$ by the amount of $\hat{x}_C$. By preparing the probe state as the maximal entangled state and performing projective measurements in the Bell basis
\begin{eqnarray}
\aligned
|\varphi_1\rangle &=\frac{|00\rangle+|11\rangle}{\sqrt{2}},\\
|\varphi_2\rangle &=\frac{|00\rangle-|11\rangle}{\sqrt{2}},\\
|\varphi_3\rangle &=\frac{|10\rangle+|01\rangle}{\sqrt{2}},\\
|\varphi_4\rangle &=\frac{|10\rangle-|01\rangle}{\sqrt{2}},\\
\endaligned
\end{eqnarray}
we get the measurement results with the probabilities given by
\begin{eqnarray}
\aligned
p_1 &=\cos^2(\tilde{B}T),\\
p_2 &=\sin^2(\tilde{B}T)\cos^2\tilde{\theta},\\
p_3 &=\sin^2(\tilde{B}T)\sin^2\tilde{\theta}\cos^2\tilde{\phi},\\
p_4 &=\sin^2(\tilde{B}T)\sin^2\tilde{\theta}\sin^2\tilde{\phi},\\
\endaligned
\end{eqnarray}
here we use $\tilde{x}=\tilde{B}(\sin\tilde{\theta}\cos\tilde{\phi},\sin\tilde{\theta}\sin\tilde{\phi},\cos\tilde{\theta})$ to denote the shifted value, i.e., $\tilde{x}=x-\hat{x}_C$. A standard procedure for the estimation of the magnetic field is to take the measurements $n$ times, and count the occurrence of each measurement results,say $k_i$, $i\in\{1,2,3,4\}$, here $n=k_1+k_2+k_3+k_4$. One then adjusts the estimation based on these occurrences: from the ratio $\frac{k_1}{k_2+k_3+k_4}$ one can adjust the estimation of $\tilde{B}$, from the ratio $\frac{k_2}{k_3+k_4}$ adjusting the estimation of $\tilde{\theta}$ and from the ratio $\frac{k_3}{k_4}$ adjusting the estimation of $\tilde{\phi}$. The data has a multinomial distribution for which one can also obtain the maximum likelihood estimation by solving the following equations\cite{machine}
\begin{eqnarray}
\label{eq:est}
\aligned
p_1&=\frac{k_1}{n},\\
p_2&=\frac{k_2}{n},\\
p_3&=\frac{k_3}{n},\\
p_4&=\frac{k_4}{n},
\endaligned
\end{eqnarray}
from the solution of the equation one can get the estimator for $\hat{\tilde{x}}(k)$, and the estimator for the true value is then given by $\hat{x}(k)=\hat{x}_C+\hat{\tilde{x}}(k)$, here $k=(k_1,k_2,k_3,k_4)$.

It is well-known that when $n$ is sufficiently large this maximum likelihood estimation saturates the Cram\'er-Rao bound\cite{machine} as long as Eq.(\ref{eq:est}) has a unique solution, i,e., when $\{\tilde{B},\tilde{\theta},\tilde{\phi}\}$ are known to belong to some intervals. For example $p_1=\frac{k_1}{n}$, which is $\cos^2(\tilde{B}T)=\frac{k_1}{n}$, has a unique solution if $\tilde{B}T\in [\frac{m}{2}\pi,\frac{m+1}{2}\pi]$ for some $m\in \mathbb{N}$, i.e., when $\tilde{B}\in \frac{1}{T}[\frac{m}{2}\pi,\frac{m+1}{2}\pi]$. Since we can choose $\hat{x}_C$ to shift the parameter, without loss of generality we can take the interval as $\tilde{B}\in \frac{1}{T}[0,\frac{1}{2}\pi]$. Similarly one can get the condition on $\tilde{\theta}$ and $\tilde{\phi}$ as $\tilde{\theta},\tilde{\phi}\in [0, \frac{\pi}{2}]$. Here the main restriction is on the condition for $\tilde{B}$ as when $T$ increases the interval gets smaller. In practice one can start with small $T$, which corresponds to a large interval for $\tilde{B}$, to first get some rough estimation, then gradually increase $T$ to get more information each round. Methods to resolve the ambiguity by choosing $T$ adaptively have also been studied previously\cite{Berry2009}. We note that the adaptive choice of $T$ is needed for both the sequential feedback and the parallel scheme, here we focus on the comparison of the two schemes in any of the round of the adaptive procedure with a chosen $T$.

We first study the precision limit under the sequential feedback scheme assuming $x$ has a prior distribution $p(x)$(based on the width of $p(x)$ the evolution time is assumed to be chosen such that the maximal-likelihood estimation works). The average mean square error is given by
\begin{equation}
E(\sum_i\delta x_i^2)=\int_x\int_k p(x)p(k|x)\sum_i[\hat{x}_i(k)-x_i]^2dkdx.
\end{equation}
When $n$ is sufficiently large, for each $x$ the maximum likelihood estimator is approximately unbiased, we thus have
\begin{equation}
\label{eq:feedback}
\sum_i\delta x_i^2\geq \frac{1}{n}Tr[J^{-1}_{FA}(x)]=\frac{1}{n}Tr[J^{-1}(x-\hat{x}_C)],
\end{equation}
here $J_{FA}(x)$ is the quantum Fisher information matrix at $x$ under the feedback scheme, which equals to the quantum Fisher information matrix(without controls) at $x-\hat{x}_C$(since the feedback controls $U_1=U_2=\cdots U_N=U^\dagger_A(\hat{x}_C,t)$ just shift the parameter by $\hat{x}_C$ when $N$ is sufficiently large).

We note that in this case the optimal measurement is independent of $x$, the quantum Fisher information matrix thus equals to the classical Fisher information matrix at all $x$. The problem in this case can be reduced to a classical estimation problem and as long as $p(x)$ is within certain interval so that Eq.(\ref{eq:est}) has a unique solution the quantum Cram\'er-Rao bound is achievable asymptotically.

For the optimal parallel scheme we use the performance at the optimal point, which is $x=0$ in this case(this can be seen from Eq.(10) in the main text), to bound the average mean square error,
\begin{equation}
\label{eq:parallelsm}
E(\sum_i\delta x_i^2)\geq \frac{1}{n}Tr[J^{-1}_{paral}(0)],
\end{equation}
which just says that the average performance is worse than the performance at the optimal point.


\begin{figure}
  \centering
  \includegraphics[width=0.6\linewidth]{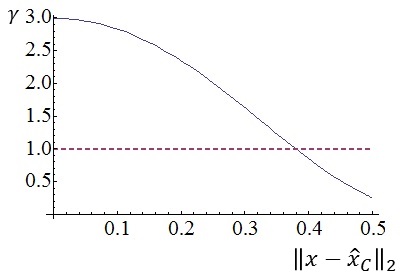}
  \caption{Comparison between quantum Fisher information matrices under the sequential feedback scheme and the optimal parallel scheme with $T=5$, $N=100$, $t=0.05$. The y-axis, $\gamma=\frac{Tr[J^{-1}_{paral}(0)]}{Tr[J^{-1}_{FA}(x-\hat{x}_C)]}$ with $J_{paral}$ and $J_{FA}$ denote the QFIMs under the optimal parallel scheme and the feedback scheme respectively; the x-axis, $\|x-\hat{x}_C\|_2$, quantifies the deviation of the estimation $\hat{x}_C$ from the true value. In this case the quantum Fisher information matrix under the feedback scheme leads to better precision than the quantum Fisher information matrix under the optimal parallel scheme as long as $\|x-\hat{x}_C\|_2\leq 0.379$. }
  \label{fig:compare}
\end{figure}
We then compare the bounds in Eq.(\ref{eq:feedback}) and Eq.(\ref{eq:parallelsm}). Under the feedback scheme from Eq.(\ref{eq:JTsm}) we have $Tr[J^{-1}_{FA}(x)]=Tr[J^{-1}(x-\hat{x}_C)]=\frac{1}{4}[\frac{1}{T^2}+\frac{2\tilde{B}^2}{\sin^2(\tilde{B}T)}]$, here $\tilde{B}=\|x-\hat{x}_C\|_2$. While under the optimal parallel scheme, from Eq.(17) of the main text we have $Tr[J^{-1}_{paral}(0)]=\frac{9}{4N(N+2)t^2}\approx \frac{9}{4T^2}$. It is then easy to see that as long as $\frac{2\tilde{B}^2}{\sin^2(\tilde{B}T)}\leq \frac{8}{T^2}$, i.e., as long as $\tilde{B}T=\|x-\hat{x}_C\|_2T\leq 1.895$, the feedback scheme outperforms the parallel scheme. Thus as long as the prior distribution, $p(x)$, is mainly concentrated in the region of $\|x-\hat{x}_C\|_2\leq \frac{1.895}{T}$, the feedback scheme outperforms the parallel scheme. This is about the same width for Eq.(\ref{eq:est}) to have a unique solution, which is about $\frac{0.5\pi}{T}\approx \frac{1.57}{T}$, thus it does not require more prior information for the feedback scheme to outperform the parallel scheme than what is already required for the maximum-likelihood estimation to work. 

 The intuition that the sequential feedback scheme can outperform the parallel scheme is because the sequential feedback scheme makes use of the information in the prior distribution by designing the feedback controls, while the parallel scheme, which undergoes the free evolution, ignores that information. The exact gain of the feedback scheme with a prior distribution should be bigger than the above analysis as the lower bound in the parallel scheme used in the above analysis is quite loose. We note that the adaptive choice of $T$ is needed for both sequential feedback scheme and the parallel scheme in order to resolve the ambiguity. We expect that the sequential feedback scheme can still outperform the parallel scheme even when the evolution time is not adaptively chosen to resolve the ambiguity, as long as the prior distribution is not close to be a uniform distribution, since some information will then be encoded in the prior distribution that can be used for the design of the feedback controls(if the prior distribution is uniform the sequential feedback scheme does not outperform the parallel scheme\cite{ChiribellaDP08} since the prior distribution contains no additional information that can be used to design the feedback controls). If the evolution time is not adaptively chosen, the performance of both schemes will deteriorate in a similar way due to the ambiguity(the parallel scheme may deteriorate faster due to the incompatibility of the optimal probe states and the optimal measurements at different values of the parameters, which however needs further investigation). Future research includes quantifying the gain of the sequential feedback scheme exactly under any prior distribution.


\end{document}